\documentclass[a4paper,fleqn,useAMS,usenatbib]{mnras}

\pdfoutput=1

\usepackage{mathptmx}
\usepackage[T1]{fontenc}
\usepackage{ae,aecompl}

\usepackage[dvips]{graphicx}
\usepackage{amsmath}
\usepackage{amsfonts}
\usepackage{amssymb}
\usepackage{comment}
\usepackage[dvipsnames]{xcolor}
\usepackage[%
        ]{hyperref}
        
\hypersetup{
	colorlinks=true,	% false: boxed links; true: colored links
	urlcolor=MidnightBlue,		% color of external links
	pdfpagelabels=true,
	hypertexnames=true,
	plainpages=false,
	naturalnames=true,
	pdftitle={Black Swans},    % title
	pdfauthor={Kipping},     % author
	linkcolor=WildStrawberry,          % color of internal links (change box color with linkbordercolor)
	citecolor=ForestGreen,        % color of links to bibliography
}

\newcommand{\pdf}{\mathrm{Pr}}

\title[Black Swans in Astronomical Data]{
Black Swans in Astronomical Data
}
\author[David Kipping]{David Kipping$^{1}$\thanks{E-mail:
\href{mailto:dkipping@astro.columbia.edu}{dkipping@astro.columbia.edu}}\\
$^{1}$Dept. of Astronomy, Columbia University, 550 W 120th Street, New York NY 10027}

\date{Accepted . Received ; in original form }

\pubyear{2020}

\begin{document}
\label{firstpage}
\pagerange{\pageref{firstpage}--\pageref{lastpage}}
\maketitle

\begin{abstract}
Astronomy has always been propelled by the discovery of new phenomena lacking
precedent, often followed by new theories to explain their existence and
properties. In the modern era of large surveys tiling the sky at ever high
precision and sampling rates, these serendipitous discoveries look set to
continue, with recent examples including Boyajian's Star, Fast Radio Bursts
and `Oumuamua. Accordingly, we here look ahead and aim to provide a statistical
framework for interpreting such events and providing guidance to future
observations, under the basic premise that the phenomenon in question
stochastically repeat at some unknown, constant rate, $\lambda$.
Specifically, expressions are derived for 1) the \textit{a-posteriori}
distribution for $\lambda$, 2) the \textit{a-posteriori}
distribution for the recurrence time, and, 3) the benefit-to-cost ratio
of further observations relative to that of the inaugural event.
Some rule-of-thumb results for each of these are found to be
1) $\lambda < \{0.7, 2.3, 4.6\}\,t_1^{-1}$ to $\{50, 90, 95\}\%$ confidence
(where $t_1=$ time to obtain the first detection), 2) the recurrence time
is $t_2 < \{1, 9, 99\}\,t_1$ to $\{50, 90, 95\}\%$ confidence, with
a lack of repetition by time $t_2$ yielding a $p$-value of $1/[1+(t_2/t_1)]$,
and, 3) follow-up for $\lesssim 10\,t_1$ is expected to
be scientifically worthwhile under an array of differing assumptions about
the object's intrinsic scientific value. We apply these methods to the Breakthrough
Listen Candidate 1 signal and tidal disruption events observed by \textit{TESS}.
\end{abstract}


\section{Introduction}
The cosmos is a vast enigmatic volume, permeated with mysteries and treasures
waiting to be known. History has demonstrated that our studies of the cosmos,
and the discoveries integral to that process, have often defied intuition
or commonly held expectation \citep{kuhn:1957}. In a Universe where our
presuppositions are often wrong, the philosophy of ``just look'', with clear
eyes and open minds, has often proven to be royal road to success 
\citep{lang:2010}.

Serendipity has played a particularly important role in this regard
\citep{fabian:2009}. Many crucial discoveries in astronomy were simply
chance accidents, with no broad expectation for their existence before-hand.
Famous examples include the Galilean moons \citep{galilei:1610}, the discovery
of Uranus \citep{herschel:1781}, radio emission from the Milky Way
\citep{jansky:1933}, cosmic X-rays sources \citep{giacconi:1962}, gamma-ray
bursts \citep{klebesadel:1973}, pulsars \citep{hewish:1968} and hot-Jupiters
\citep{mayor:1995}.

In most cases, these detections were found to represent the tip of a proverbial
iceberg; the inaugural member of a new class of astronomical phenomena. For
example, hot-Jupiters are now known to reside around 0.4\% of FGK-dwarf stars
with new discoveries being somewhat routine today \citep{zhou:2019}. And yet,
those ``firsts'' are perhaps the moments of greatest excitement, when observers
naturally wonder about how many analogs might exist, and when the next one
will be found. For example, shortly after the discovery of the first
interstellar asteroid, `Oumuamua \citep{meech:2017}, estimates of the
implied number density soon followed \citep{trilling:2017,do:2018} - estimates
which were of course conditioned upon a single data point\footnote{
Other recent examples include Boyajian's Star \citep{boyajian:2016}
and the Lorimer burst \citep{lorimer:2007}.
}. It is sometimes
said that ``a single data point teaches you nothing'', but truly that
describes zero data points. Inference \textit{can} proceed with a single datum
(and evidently do), but naturally the associated uncertainty of such inference
will be greater than that conditioned upon $>1$ data points.

The situation is somewhat analogous to trying to infer the rate of black swans
that one expects to observe having thus far seen a singular example. And so, by
metaphor, a ``black swan'' event often refers to an unanticipated, rare event
of particular importance or significance \citep{taleb:2010}.

In astronomy, the rate calculation of a ``black swan'' event, like `Oumuamua,
can take different forms. In \citet{trilling:2017} and \citet{do:2018}, the
authors are principally interested in the true number density of such objects,
which requires a careful treatment of the observational selection biases. This
naturally depends upon the observing strategy and sensitivity of the telescopes
involved and will thus be generally bespoke to that particularly phenomenon.
However, a more general case is apparent when one asks - what is the rate at
which one should expect to observe these events if my observations continue
without any modification? This sidesteps the issue of selection bias since
one is not asking about the intrinsic rate, but the \textit{observed} rate.
Similarly, one can ask questions building off this, such as when should I
expect to see a recurrence and how long is it worth my time to observe waiting
for such analogs?

In this work, a general statistical framework for tackling these questions is
sought. The underlying assumption, as already stated, is that the
observational selection biases do not change. If they do, this would clearly
need to be accounted for. Whilst the idea of a fixed, unchanging observational
mode might initially seem somewhat unrealistic or uncommon, it is suggested
here that this is indeed becoming much more prevalent with new and upcoming
missions. In the modern era of all-sky astronomical surveys, such as
\textit{Gaia} \citep{gaia:2016}, \textit{TESS} \citep{tess:2015} and the Vera
Rubin Observatory (VRO; \citealt{VRO:2019}), the observing strategy and
sensitivity are often pre-decided and fixed for many years at a time. In this
way, fixed observational biases are becoming common, indeed highly desirable
as a means to more straight-forwardly invert detection rates into occurrence
statistics. Further, the enormous volumes of data being produced by these
surveys (e.g. \citealt{juric:2017}) has demoted the role of human ``by eye''
detections (which certainly suffers from time/caffeine dependent sensitivity),
such that increasingly homogeneous, automated machine learning approaches are
being deployed to detect astronomical phenomena (e.g. \citealt{wagstaff:2016,
gonzalez:2018,lin:2020}), including anomalies (e.g. \citealt{giles:2019,
wheeler:2019,storeyfisher:2020}).

This paper is organized as follows. In Section~\ref{sec:post}, the
\textit{a-posteriori} distribution for the observed anomaly rate is
derived, and related summary statistics. In Section~\ref{sec:recurrence},
the time until the next recurrence is derived in a probabilistic sense,
with implications for applying tension to the hypothesis of repeatability.
In Section~\ref{sec:BCR}, a simple cost-benefit analysis is provided as a
way of guiding the economical use of continued time/effort to seek
analogs of the black swan event.

\section{Posterior Distribution for the Anomaly Rate}
\label{sec:post}

\subsection{Formalism}

Consider observing an object, or a sample of objects, with time series
astronomical observations. Let us further consider the simplifying case
where the data are homogenous, regular and no substantial changes
to the observing instruments nor strategy are implemented. After a time
$t_1$, a highly significant anomaly is detected for which there is no
precedent and is confirmed to be of astrophysical origin. If the anomaly
is short-lived and transient, it may be impossible to secure any further
data about the anomaly, besides the fact it occurred.

In such a scenario, one faces the frustrating situation of seeing something
truly remarkable yet for which there is essentially no way of obtaining
any further information. Naturally, astronomers may at this point wonder
how many analogs of this anomaly exist, and when the next one might be
expected to be found. And yet, astronomers are forced to make
inferences conditioned upon this sole black swan event.

To make mathematical progress, let us assume that the anomaly (or an event
sufficiently similar to be classified as analogous), will indeed repeat at
some future time, or within some similar sample, given sufficient observations.
Let us further assume that the intrinsic rate at which this anomaly appears
in our observations is constant over time. In other words, we do not live
at a particularly ``special'' time (or equivalently did observe a
particularly special sample) and thus the anomaly rate is uniform, such
that the probability of detection is the same in any given equal duration
time interval (or equivalently sample size).

The above describes a Poisson process and hence provides the basis of making
analytic progress. In some cases, a Poisson process is not directly
appropriate (e.g. repeating FRBs from an individual source can be non-Poisson;
\citealt{oppermann:2018}) but can often be made so by simple re-framing of the
problem (e.g. the number of FRBs received over the entire sky from
independent sources over a given interval will be Poissonian).
For a Poisson process, the probability of observing $N$ events
over a time interval $t$ is given by

\begin{align}
\pdf(N|\lambda,t) &= \frac{1}{N!} e^{-\lambda t} (\lambda t)^N,
\end{align}

where $\lambda$ is the intrinsic rate per unit time of the process. Note that
our formalism is expressed in terms of time, but could equally be replaced
with data volume, sample size or even ``effort level'' more generally. However,
time provides a clarifying example and thus we continue with this notation
in what follows.

When the first event is detected, all that is known is how long it took to
obtain that first detection. We consider that ``now'' represents the moment
of the anomaly detection and impatient astronomers across the world are
immediately wondering when the next event will recur. Thus, the only data
in-hand is that it took a time $t_1$ to obtain that one event.

A well-known result of Poisson processes is that the time interval between
successes follows an exponential distribution \citep{cooper:2005}.
Equivalently, it must hold that the probability distribution for the time it
takes to obtain our first success when initiating from arbitrary reference time
is also an exponential, since a Poisson process is - by definition - an
independent process with no ``memory'' of previous events, much like how your
chance of rolling a six on a dice has no dependency on previous rolls.
Accordingly, the time for a first success follows

\begin{align}
\pdf(t_1|\lambda) &= \lambda e^{-\lambda t_1}.
\label{eqn:t1}
\end{align}

%A simple derivation is to consider that is that
%
%\begin{align}
%\pdf(N>0|\lambda,t) &= 1 - \pdf(N=0|\lambda,t),\nonumber\\
%\qquad&= 1 - e^{-\lambda t},
%\end{align}
%
%and since this is the probability of obtaining at least one success over a time
%interval $t$ (i.e. a cumulative probability) then the probability distribution
%of $t$ itself must be the temporal derivative of this (giving
%Equation~\ref{eqn:t1}).

\subsection{An \textit{a-posteriori} distribution for the anomaly rate, $\lambda$}

To infer $\lambda$ from $t_1$, one can use Bayes' theorem \citep{bayes:1763} to
write that

\begin{align}
\pdf(\lambda|t_1) \propto \pdf(t_1|\lambda) \pdf(\lambda).
\end{align}

\begin{figure*}
\begin{center}
\includegraphics[width=15.0cm,angle=0,clip=true]{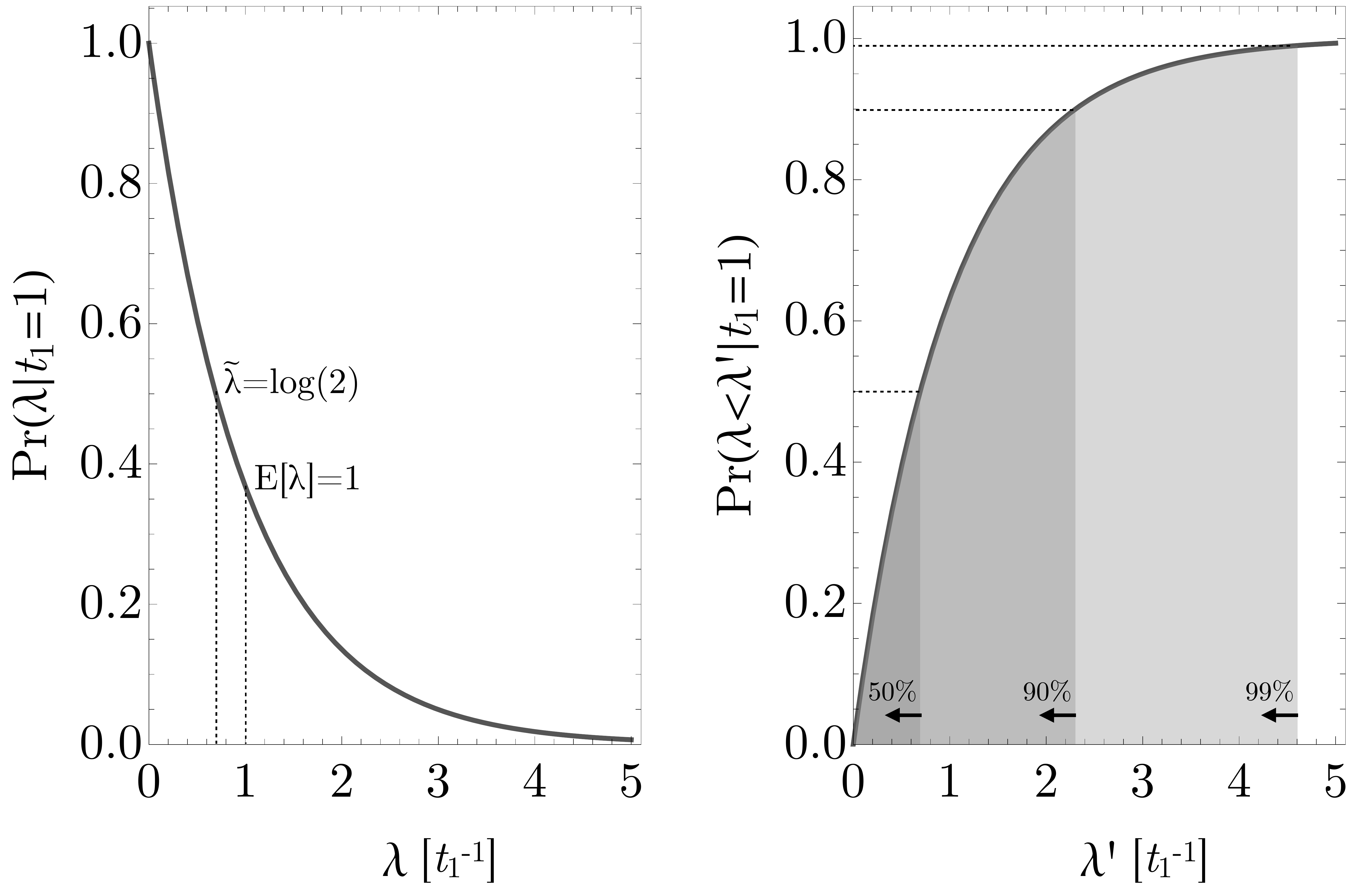}
\caption{
\textbf{Left:} The \textit{a-posteriori} distribution of $\lambda$, the anomaly
rate, given one detection of the anomaly in hand. \textbf{Right:} The cumulative
density function of the posterior highlighting three useful upper limits.
}
\label{fig:post_dist}
\end{center}
\end{figure*}

For $\pdf(\lambda)$, the prior, a scale-invariant objectively defined
distribution in sought - which is provided by using the approach of
\citet{jeffreys:1946}. Here, the Fisher information \citep{fisher:1922} of an
exponential distribution is $\lambda^{-2}$ and thus the Jeffreys prior is
simply $\lambda^{-1}$. After normalizing the posterior, one obtains

\begin{align}
\pdf(\lambda|t_1) &= t_1 e^{-\lambda t_1},
\label{eqn:lambdapost}
\end{align}

whose behaviour is depicted in Figure~\ref{fig:post_dist}. Note
that this should not be applied to the rate of abiogenesis given the timing of
the first appearance of life, since our existence is predicated upon a
success and thus is severely sculpted by selection bias \citep{spiegel:2012}.
In contrast, it assumed here that our existence is no way dependent upon the
anomaly being detected, which of course can be safely assumed in almost all
other situations.

\subsection{Properties of the posterior distribution}

Some basic properties of this distribution are that it has a mode at
$\lambda=0$ and monotonically declines out to infinity, with an expectation
value of $1/t_1$, a median of $(\log 2)/t_1$ and a variance of $1/t_1^2$. If one
works in units of time such that $t_1=1$, then one can express that
$\tau\equiv1/\lambda$ (the characteristic timescale) is $1\pm1$ (for the mean
and standard error).

Since the function is monotonically decreasing from $\lambda=0$, one can place
upper limits on $\lambda$ as $\lambda < -\log(1-C)$ to a confidence level of
$C$. For example, $\lambda < \{0.693, 2.302, 4.605\}\,t_1^{-1}$ to $\{50, 90,
99\}$\% confidence. This can be trivially inverted to $\tau$, however, in doing
so one obtains \textit{lower} limits on the timescale (rather than upper limits
on the rate). Specifically, one finds $\tau > \{1.443, 0.433, 0.217\}\,t_1$ to
$\{50, 90, 99\}$\% confidence (or more generally $\tau/t_1 > -1/\log(1-C)$ to a
confidence level $C$).

\section{Predicting the Next Recurrence}
\label{sec:recurrence}

\subsection{Recurrence time posterior}

Having observed the black swan anomaly, and characterized the probability
distribution of its rate, $\lambda$, our next question might be when should
one expect its recurrence? Let us define that the recurrence time (as measured
since time $t_1$) is given by $t_2$. Recall that the time until the next event
must follow an exponential distribution, and so the distribution of $t_2$ is

\begin{align}
\pdf(t_2 | \lambda) &= \lambda e^{-\lambda t_2}.
\end{align}

\begin{figure*}
\begin{center}
\includegraphics[width=15.0cm,angle=0,clip=true]{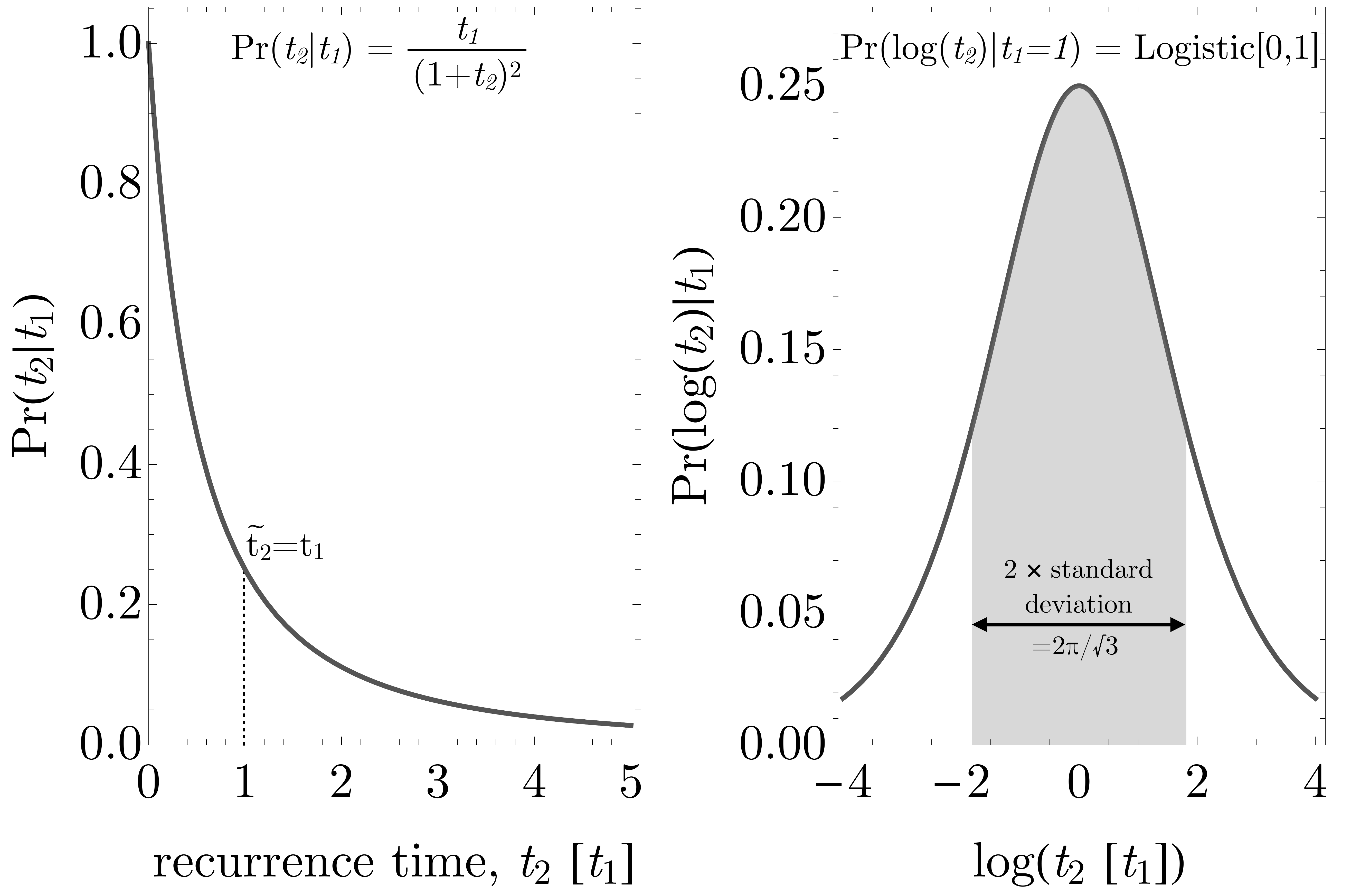}
\caption{
\textbf{Left:} The \textit{a-posteriori} distribution of the anomaly recurrence
time, $t_2$, given one detection of the anomaly in hand. \textbf{Right:}
Same as left, except we show the distribution of the logarithm of
$t_2$.
}
\label{fig:rec_dist}
\end{center}
\end{figure*}

Critically, $\lambda$ in the above is now ``known'', at least in a
probabilistic sense by virtue of our posterior distribution in
Equation~(\ref{eqn:lambdapost}). Accordingly, the distribution of
$t_2$, accounting for our constraints on $\lambda$, can be found through
marginalization:

\begin{align}
\pdf(t_2 | t_1 ) &= \int_{\lambda=0}^{\infty} \pdf(t_2 | \lambda) \pdf(\lambda | t_1) \mathrm{d}\lambda,\nonumber\\
\qquad&= \frac{1}{ (1 + t_2)^2 },
\label{eqn:t2dist}
\end{align}

whose behaviour is illustrated in Figure~\ref{fig:rec_dist}.

\subsection{Logarithm of the recurrence time}

Another useful way to think about the above is in logarithmic time,
where one can transform the above to find that $\log t_2$
follows a logistic distribution with a mean of zero and
scale-parameter unity:

\begin{align}
\pdf(\log(t_2) | t_1 ) &= \frac{ e^{-\log t_2} }{ (1 + e^{-\log t_2})^2 },
\end{align}

This is convenient because a logistic distribution is quasi-Gaussian
\citep{gauss:1809} in shape (see Figure~\ref{fig:rec_dist}), and thus one can
approximate that the logarithm of the time until the next event is
approximately Gaussian with a mean of $0$ and standard deviation
$\pi/\sqrt{3} \simeq 1.8$ (recall we are working in units of $t_1$ time).
In other words, the recurrence time is approximately log-normal (and correctly
log-logistic).

\subsection{Upper limits on the recurrence time}
\label{sub:recurrencelimits}

Coming back to linear space (Equation~\ref{eqn:t2dist}), the mean of the
distribution is infinite and doesn't present a useful summary statistic (and
thus so too is the variance). As with the $\lambda$ posterior, $t_2$ is
found to follow a monotonically decreasing function peaking at zero. The
cumulative distribution is analytic (simply $t_2/(1+t_2)$) and
thus provides useful quantiles. This allows to state a lower limit that
$t_2/t_1 < C/(1-C)$, to a confidence level of $C$. To give examples,
$t_2/t_1 < \{1, 9, 99\}$ to $\{50, 90, 99\}$\% confidence level.

An important consequence of the above is that even after waiting 9 times
longer than the initial detection time, there is still a respectable
10\% probability that one will not have observed a recurrence yet. Black
swans demand patience.

These results can also be used in testing the hypothesis that the event
is indeed repeating. Whilst the black swan event is assumed to be
astrophysical nature here with a uniform rate, it could instead by a
non-repeating event caused by the instrument itself, for example. In
such a scenario, after observing the event for $t_2 = 99 t_1$,
statistical pressure would bear down on the hypothesis, essentially
producing a p-value of 1\%. Going further, we can write that the
$p$-value for this hypothesis is $1/[1+(t_2/t_1)]$. Rigorous model
testing might proceed then through comparison to a suitable instrument
outlier model, as $p$-values alone are fraught as a sole means of
hypothesis exclusion \citep{wasserstein:2016}.

\subsection{Constraints from future null detections}

Suppose observations of the anomaly's source continue after its initial
detection. Let us assume that these subsequent observations span a time
interval $t_{\mathrm{obs}}$, but one finds no further anomalies. How does this
new information - a lack of subsequent detections - affect our inference on
$\lambda$?

The probability of observing zero events over a time interval
$t_{\mathrm{obs}}$ can be evaluated from the Poisson distribution

\begin{align}
\pdf(N=0;\lambda,t_{\mathrm{obs}}) &= e^{-\lambda t_{\mathrm{obs}}}.
\end{align}

This forms a monotonically decreasing exponential curve, where for
any given choice of rate, one can see how ``dry spells'' much longer than
than $1/\lambda$ are improbable - matching our intuition for the
problem. And so one may employ this as a likelihood function for
$\pdf(t_{\mathrm{obs}}|\lambda)$ and then via Bayes' theorem, write

\begin{align}
\pdf(\lambda|t_{\mathrm{obs}}) \propto \pdf(\lambda|t_{\mathrm{obs}})\pdf(\lambda).
\end{align}

One can go further by folding in the information already learned about
$\lambda$ via the prior:

\begin{align}
\pdf(\lambda|t_{\mathrm{obs}},t_1) \propto \pdf(\lambda|t_{\mathrm{obs}})\pdf(\lambda|t_1),\nonumber\\
\pdf(\lambda|t_{\mathrm{obs}},t_1) \propto e^{-\lambda t_{\mathrm{obs}}} (t_1 e^{-\lambda t_1}).
\end{align}

After normalization, this becomes

\begin{align}
\pdf(\lambda|t_{\mathrm{obs}},t_1) &= (t_1+ t_{\mathrm{obs}}) e^{-\lambda (t_1+ t_{\mathrm{obs}})}.
\label{eqn:lambdapost2}
\end{align}

By comparison to $\pdf(\lambda|t_1)$ in Equation~(\ref{eqn:lambdapost2}),
one can see that this is equivalent to simply setting $t_1 \to (t_1 +
t_{\mathrm{obs}})$. This make sense since the Poisson process is independent
and thus doesn't ``know'' that we stopped the clock at $t_1$ before, only that
1 event occurred over said time interval.

\section{Cost-Benefit Considerations of Future Observations}
\label{sec:BCR}

\subsection{Cost-benefit analysis of null results}
\label{sub:BCR1}

As shown earlier in Section~\ref{sub:recurrencelimits}, there is a 50\% chance
of a recurrence by observing over the same time interval as for the first
anomaly's identification. But, going for twice that time only increases the
odds to 67\% ($+17$\%), and thrice that to 75\% ($+8$\%). This implies that the
initial observations after the event are worthwhile, but the longer one
observes, the slower one's odds of success grow. In models of diminishing
returns, especially those with associated costs, one might anticipate an
optimal observing strategy to emerge.

Continuous observations can improve our knowledge in two basic ways. First, a
lack of any positive detections is useful in that it constrains the rate
parameter, $\lambda$. Here, one is really interested in how much tighter the
\textit{a-posteriori} distribution of $\lambda$ becomes when conditioned upon
additional data featuring no detections. Second, a success will, of course,
offer the opportunity to not only constrain $\lambda$ much better, but also
provide additional objects for which their character and properties can be
compared to the original. The ``value'' of that second possibility is difficult
to directly compare to the act of improving our knowledge of $\lambda$, and so
it is treated here as a separate problem (see Section~\ref{sub:BCR2}). But in
both cases, one can consider a simple cost-benefit optimization to provide
guidance to observers.

Let us first consider the possibility of no new detections after observing a
time interval $t_{\mathrm{obs}}$ after the first detection. This act naturally
improves our constraint on $\lambda$ and one can characterize
the ``value'' associated with that improvement by evaluating the information
gained using information theory. This is quantified by the Kullback-Leiber
Divergence (KLD; \citealt{KLD:1951}) going from $\pdf(\lambda|t_1) \to
\pdf(\lambda|t_2,t_1)$, which evaluates the change in Shannon entropy
\citep{shannon:1948} in expending the additional observing time
$t_{\mathrm{obs}}$, and is calculated as

\begin{align}
\mathrm{KLD} &= \int_{\lambda=0}^{\infty} \pdf(\lambda|t_{\mathrm{obs}},t_1) \log \Big(
\frac{\pdf(\lambda|t_{\mathrm{obs}},t_1)}{\pdf(\lambda|t_1)}
\Big) \mathrm{d}\lambda,\nonumber\\
\qquad&= \frac{ (t_1 + t_{\mathrm{obs}})(\log(t_1 + t_{\mathrm{obs}}) - \log t_1) - t_{\mathrm{obs}} }{t_1 + t_{\mathrm{obs}}}\,\mathrm{nats}.
\end{align}

\begin{figure}
\begin{center}
\includegraphics[width=8.4cm,angle=0,clip=true]{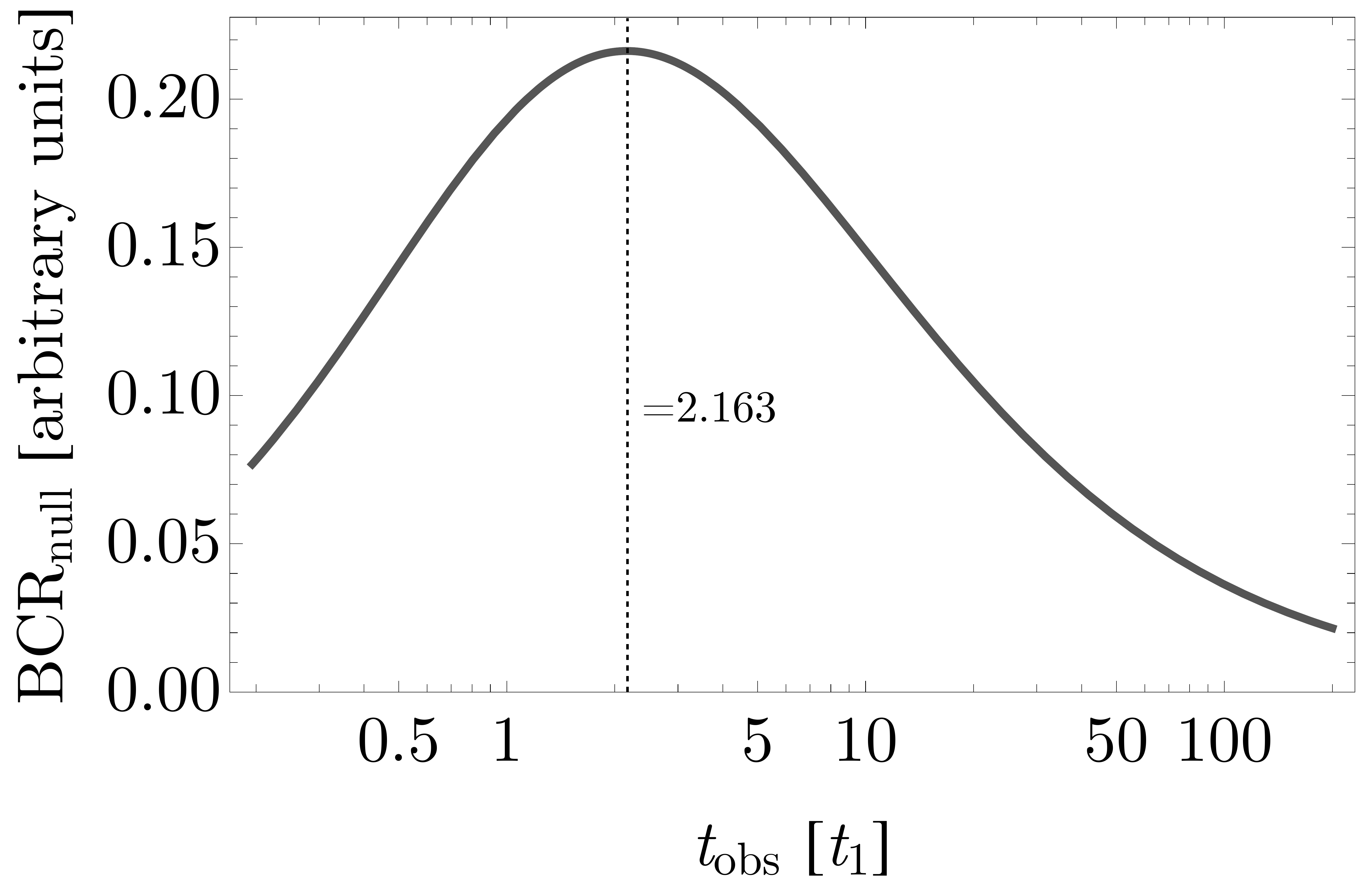}
\caption{
Benefit to cost ratio (BCR) obtained from a continuous null detections of
a repeat of the black swan anomaly, as characterized by the improvement
in the underlying anomaly rate. A single peak occurs at just over twice
the initial time to obtain a detection, after which point observations
become less economical.
}
\label{fig:KLD}
\end{center}
\end{figure}

Information always increases with more time, but of course in many applications
that time, or more generally ``effort level'', comes at cost. Assuming cost
scaling proportional to time, then the information/time provides a direct way
of optimizing resources in a cost-benefit analysis - at least under the defined
objective of trying to better constrain $\lambda$.
Let us define that the benefit-to-cost ratio of obtaining further
null results, $\mathrm{BCR}_{\mathrm{null}}$, is proportional to the
KLD in the above divided by the time used $t_{\mathrm{obs}}$,
and further work in units of time $t_1$ to yield

\begin{align}
\mathrm{BCR}_{\mathrm{null}} &\propto \frac{ (1+t_{\mathrm{obs}}) \log(1+t_{\mathrm{obs}}) - t_{\mathrm{obs}}}{t_{\mathrm{obs}}(1+t_{\mathrm{obs}})},
\end{align}

which is plotted in Figure~\ref{fig:KLD}. In this case, we note that
the cost-to-benefit ratio has a single maximum
located at $t_{\mathrm{obs}} = 2.174 t_1$.

This reveal, then, that if one has observed the source for ${\sim}$twice as
long as the initial run and still not seen a recurrence, it's becoming
uneconomical to continue - in the sense of a null-result providing worthwhile
learning on $\lambda$. Of course, this does not account for the value of a
possible detection, which is addressed next.

\subsection{Cost-benefit analysis of a fishing expedition}
\label{sub:BCR2}

Let's assume that the first anomaly detection has a ``value'' $v_1$.
Scientific value is somewhat subjective and field/goal specific,
but could, for example, be a proxy for scientific impact (such
as citation count). Since it
took a time $t_1$ to obtain that success, then the benefit-to-cost ratio of
the first anomaly is simply $\mathrm{BCR}_1 \propto v_1/t_1$. Subsequent
successes may not be necessarily the same value as the first, and indeed will
typically be less scientifically valuable. For the moment, we leave the
functional form for these subsequent values undefined, but adopt the notation
that the value of the $n^{\mathrm{th}}$ anomaly is $v_n$. Accordingly, the
total value of finding $m$ \textit{new} objects (hence $m+1$ objects in total
now known) will be

\begin{align}
V_m &= \sum_{n=2}^{m+1} v_n.
\label{eqn:sum}
\end{align}

Consider at time $t_1$, after the announcement of the anomaly, one is planning
a new observing program. Consider further that for the new observing plan, a
time interval $t_{\mathrm{obs}}$ is selected and fixed. The expected
total value, $<V_{\mathrm{tot}}>$, of this observing run is found by summing
the total value of $m$ successes with the probability of $m$ successes over
all $m$ indices (essentially a weighted average):

\begin{align}
<V_{\mathrm{tot}}> &= \sum_{m=1}^{\infty} V_m \Big(\frac{ e^{-\lambda t_{\mathrm{obs}}} (\lambda t_{\mathrm{obs}})^m }{ m! }\Big).
\end{align}

Note that the above depends on $\lambda$, for which we have a posterior
(Equation~\ref{eqn:lambdapost}). One can thus include this via marginalization,
as before, where we use the bar symbol above our expectation value to denote
that it is marginalization:

\begin{align}
<\bar{V_{\mathrm{tot}}}> &= \int_{\lambda=0}^{\infty} \sum_{m=1}^{\infty} V_m \Big(\frac{ e^{-\lambda \Delta_{\mathrm{obs}}} (\lambda \Delta_{\mathrm{obs}})^m }{ m! }\Big) t_1 e^{-\lambda t_1} \,\mathrm{d}\lambda.
\end{align}

To make progress, one can pull the sum outside, switch to units of time
relative to $t_1$, and evaluate the individual integrals as

\begin{align}
<\bar{V_{\mathrm{tot}}}> &= \sum_{m=0}^{\infty} \int_{\lambda=0}^{\infty} V_m \Big(\frac{ e^{-\lambda t_{\mathrm{obs}}} (\lambda t_{\mathrm{obs}})^m }{ m! }\Big) e^{-\lambda} \,\mathrm{d}\lambda\,\nonumber\\
\qquad&= \sum_{m=0}^{\infty} V_m \Big(\frac{t_{\mathrm{obs}}^{m}}{(1+t_{\mathrm{obs}})^{m+1}}\Big) \Big(\frac{\Gamma[m+1]}{m!}\Big).
\end{align}

Finally, we normalize with respect to the cost by dividing by
$t_{\mathrm{obs}}$. This yields a monotonically decreasing function
that peaks at zero with a value of $\mathrm{BCR}_1$:

\begin{align}
\frac{<\mathrm{BCR}>}{\mathrm{BCR}_1} &= \sum_{m=0}^{\infty} \Big(\frac{V_m}{v_1}\Big) \Big(\frac{t_{\mathrm{obs}}^{m-1}}{(1+t_{\mathrm{obs}})^{m+1}}\Big) \Big(\frac{\Gamma[m+1]}{m!}\Big).
\label{eqn:BCR2}
\end{align}

We consider three different toy models of valuing the $n^{\mathrm{th}}$
detection. We caution against taking these models too seriously; they are
here to guide our intuition and reveal the functional properties in this
analysis. The first is a simple power-law of the form $v_n = v_1 n^{-\alpha}$
which implies, via Equation~(\ref{eqn:sum}), that $V_m = v_1
(\mathrm{H}_{m+1}^{(\alpha)} - 1)$ where $\mathrm{H}_{l}^{(\alpha)}$ is the
$l^{\mathrm{th}}$ Harmonic number of order $\alpha$. A typical choice might be
$\alpha = 1/2$, where the second detection is valued at $1/\sqrt{2}$ of the
first for example. Another reasonable choice might be $\alpha=1$, which has
imbues a quicker decline in value.

The second value model we consider starts from the philosophy of the summed
values. If one has a sample of $l$ objects in hand, and one attempts to
measure the mean of some property of this sample, our uncertainty on the mean
would be proportional to $1/\sqrt{l}$. Let us define our value to be inversely
proportional to the error on the mean, which means $v_n = v_1 \sqrt{n}$,
and summing with Equation~(\ref{eqn:sum}), one finds $V_m = v_1(\sqrt{m+1}-1)$.

The third and final value model considered is similar to the second, except
we refine the assumption that value is inversely proportional to error on the
mean. Instead, this is replaced with the KLD between two Gaussians centered on
the same mean, with a standard deviation going from $\sigma\to\sigma/\sqrt{l}$
for a collection of $l$ objects. This essentially means that our value
is directly proportional to the information gained about the properties of
the anomaly. Accordingly, one finds $v_n = v_1 (\log(\sqrt{n}) -
\log(\sqrt{n-1}))$ and thus $V_m = v_1 \log(m+1)$.

\begin{figure*}
\begin{center}
\includegraphics[width=17.4cm,angle=0,clip=true]{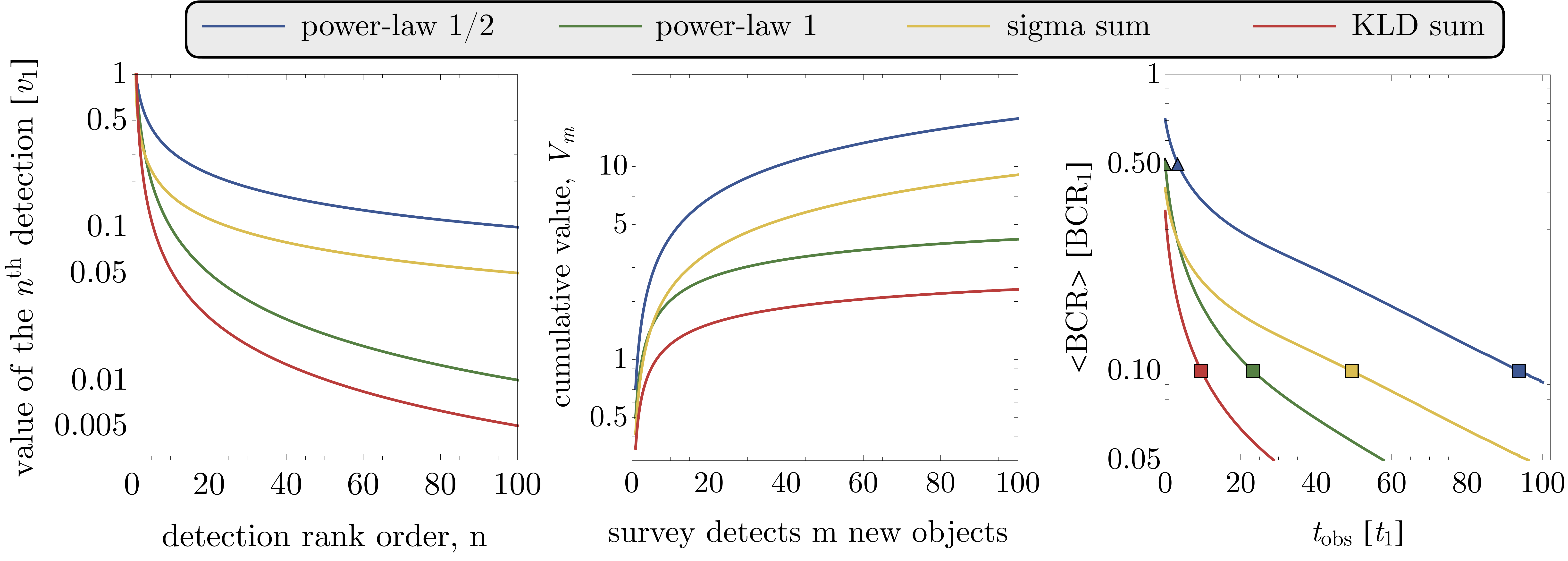}
\caption{
\textbf{Left:} 3+1 toy models (two of them are both power laws) for valuing
the scientific value of subsequent detections of a black swan anomaly.
\textbf{Middle:} Cumulative sum of the valuations from the 2nd to the
$(m+1)^{\mathrm{th}}$ detection.
\textbf{Right:} Expected benefit-to-cost ratio of observing for a time
$t_{\mathrm{obs}}$ when subsequent anomalies are valued using the
models depicted in panels 1 and 2.
}
\label{fig:grid}
\end{center}
\end{figure*}

The value of each detection $v_n$, and the summed values $V_m$, are plotted
in Figure~\ref{fig:grid}. This reveals that the third valuation system,
based on the KLD, yields the most pessimistic valuation over the range
shown, where as the power-law of index $\alpha=1/2$ is the most optimistic.
Although we cannot analytically evaluate Equation~(\ref{eqn:BCR2}) with these
three functions, they are numerically evaluated them up to
$t_{\mathrm{obs}} = 10 t_1$ for $m=10^4$ terms.

All of the functions are monotonically decreasing with $t_{\mathrm{obs}}$ and
peak at $t_{\mathrm{obs}}=0$. However, this does not mean the efforts are
``unprofitable'', merely that they are not expected to be \textit{as}
profitable as the initial detection. For example, if the initial run was a
Nobel prize winning discovery, the scientific value would be presumed to be
extremely high and thus subsequent efforts would still be very scientifically
profitable. In this vein, let us assume that $\mathrm{BCR}_1 \gg 1$ (else
ultimately it would probably not be considered interesting enough to warrant
follow-up anyway). Let us quantify that $\mathrm{BRC}_1 > 10$ to take this as
one order-of-magnitude, then. Accordingly, any campaigns with
$\mathrm{BCR}/\mathrm{BCR}_1 > 0.1$ would be deemed profitable
and worthwhile. For the 3+1 models shown in Figure~\ref{fig:grid},
this occurs at $\sim{100} t_1$ for the most optimistic valuation model,
the power-law with $\alpha=1/2$, and at $\sim{10} t_1$ for the most
conservative valuation system, the KLD summation model.

From this, it is concluded that follow-up efforts of a black swan event
are expected to remain scientifically worthwhile and ``profitable''
for the observers for time intervals/sample sizes/effort levels
of $\lesssim 10$ times that of the original discovery, but may become
unprofitable for longer times.

\section{Two Simple Applications}
\label{sec:applications}

\subsection{Case 1: A Current Black Swan}

In order to show the derived results in action, we here present two simple
applications. In the first case, we take an example of a recent astronomy
detection which is currently considered a genuine black swan with no
subsequent repetitions. In particular, we take the example of the
detection of Breakthrough Listen Candidate 1 (BLC1) - a narrow bandwidth,
high significance radio emission apparently originating from the direction
of Proxima Centauri (Breakthrough Listen team, \textit{in prep}).

BLC1 was detected by the Breakthrough Listen observational survey operated
with the 64\,m Parkes radio telescope \citep{price:2018,lebofsky:2019,price:2020}. The signal was found
in archival observations from April to May 2019 during a focussed campaign on
the source Proxima Centauri. The signal is likely some form of human radio
transmission\footnote{See \href{https://sites.psu.edu/astrowright/2020/12/20/blc1-a-candidate-signal-around-proxima/}{this URL}},
but could, in principle, originate from alien technology in the Proxima
vicinity.

Investigation of the nature and origin of the signal continues (Breakthrough
Listen team, \textit{in prep}), but in what follows we proceed under the
assumption it is a) of astrophysical origin, and b) repeats (potentially
stochastically) over sufficiently long observing times. Ultimately, our
formalism provides a way to ask what kind of observations would put tension on
this hypothesis, and thus a pathway to eliminating the candidacy of this signal
independent of specific tests, traits or properties of the one recorded signal.

The frequency of BLC1 (982\,MHz) is such that it was only detected in one of
the four receivers used by the project, namely the ``Ultra Wideband Low''
(UWL). Proxima Centauri was observed for a total of 1550\,minutes over an
interval of 1.75\,years with UWL up to the point of its detection (A. Siemion,
priv. comm.). How long it has been observed since then is fluid, and thus we
will adopt these numbers for the purposes of this calculation.

From Equation~\ref{eqn:t2dist}, the recurrence time of BLC1 is expected to
occur by a time $t_2$, where $t_2$ follows a probability distribution given by
$[1+(t_2/t_1)]^{-2}$. Consider that the source continues to be monitored without
repetition. To 95\% confidence, we would expect this repetition to occur after
a cumulative observing time of 29,450\,minutes, or 19 times the original
observing duration. This corresponds to 20.5\,days of on-target observations.

The above considers only the on-target observing time. In practice, observing
Proxima for 20.5 days could still miss a repeating signal if the signal's
characteristic repetition time is $\gg 1550$\,minutes ${\sim1}$\,day. To
account for this possibility, the wall-time of the 29,450\,minutes (or more)
of on-target observations would need to be spread over 19 times longer that
the original wall-time, which is 33 years. Accordingly, a short-term repeating
signal from Proxima could be excluded to 95\% confidence with
${\gtrsim}$20\,days of continuous monitoring, but a long-term repeating signal
could not excluded until $\gtrsim{33}$\,years of observations.

We note that our benefit-to-cost calculations do not apply here since the
validity of the first signal is ambiguous, and thus the second signal would
arguably have greater value by virtue of confirming the nature of the signal.

\subsection{Case 2: A Once Black Swan}

Let us to turn to a case where repetition was observed and see how well
the equations presented here perform. Whilst there are many historical examples
of black swans later found to repeat, we here try to identify a case where the
signal was identified and repeated within a homogeneous, single astronomical
survey. As noted earlier, this is an inherent assumption to our formalism, in
order neatly sidestep the issues of differing observational biases between
surveys. This requirement eliminates most historical examples, but
as suggested earlier, the era of large astronomical surveys such as Vera
Rubin and the Roman telescope look primed to change this.

To this end, we take tidal disruption events (TDEs) observed photometrically by
\textit{TESS}. \textit{TESS} pursues a systematic survey of the sky observing
approximately the same number of sources at any given time \citep{tess:2015},
for which the distribution of apparent magnitudes and colors is broadly
consistent \citep{tic}. ASASSN-19bt represents the first TDE observed by
\textit{TESS}, observed simultaneously by the ground-based ASASSN survey
\citep{holoien:2019}. The TDE was observed in sectors 7, 8 and 9 peaking at
MJD\,58550, and beginning to rise 40\,days prior. Since \textit{TESS}
observations began MJD\,58324, then here we here have $t_1 = 186$\,days - as
defined by the start of the slow rise.

Let us ask, when should we expect \textit{TESS} to observe a second such TDE
somewhere on the sky, based on the first. We restrict ourselves to
non-repeating TDEs here, since repeating ones (e.g. \citealt{payne:2020}) could
clearly lead to many such events from the same source - rather than really
probing the rate at which the sky produces TDEs as seen by \textit{TESS} (as
well as representing an astrophysically distinct scenario). Accordingly, the
second TDE observed by \textit{TESS} was recently reported by
\citet{hinkle:2020}, ASASSN-20hx. This second TDE was identified in its slow
rise phase at MJD 59040 (the peak intensity time has not yet been reported),
corresponding to a time \textit{TESS}' mission of 716\,days, or
$t_2 = 530$\,days (2.85 times $t_1$).

Although $t_2$ is clearly longer than $t_1$ by nearly a factor of 3, using
the expressions of Section~\ref{sub:recurrencelimits}, the $p$-value of this
occurring is 26\% and thus hardly surprising. Indeed, the timing is fully
consistent with the expectation of the sky producing a Poisson background
rate of such TDEs. Further, we can evaluate that the expectation value
for the benefit-to-cost ratio at time $t_2$ is 0.518, 0.288, 0.286
and 0.181 (in units of the BCR of the first detection) for the power-law
1/2, power-law 1, sigma sum and KLD sum methods, respectively. Thus,
by all methods, a continued search for such events in the \textit{TESS}
was very much worthwhile ($\gtrsim 0.1$) by the point of the success.

\section{Discussion}
\label{sec:discussion}

In this work, we have considered the statistical implications of the detection
of a black swan astronomical event. Although the work is framed largely
in terms of time, where the event occurs at some fixed temporal rate,
it can be equally applied to other sample types too (e.g. objects surveyed,
frequencies scanned, wavelengths monitored), or indeed most generally
``effort level'', provided that the observational selection biases are
homogeneous across these samples, however they are defined.

It is envisioned that this work will be beneficial in the future analysis of
large astronomical survey data. Here, the assumption of unchanging
observational bias is most likely to hold, as surveys may adopt fixed
strategies for multi-year periods. However, if the black swan event is low
signal-to-noise, and was somehow fortunate to have been detected, then software
improvements in search strategy would invalidate this assumption. However,
large signal-to-noise events will likely not benefit noticeably from software
improvements. Furthermore, if they were identifiable in the first place amongst a
vast volume of data (as expected from upcoming surveys; \citealt{juric:2017}
then they were likely identified with an automated process already, and so the
assumption of constant selection effect is sound.

Implications of this work are that black swan events can have unintuitively
long recurrence times. Whilst there is a 50\% chance of the event repeating
within one more unit of effort, 99 times is needed to reach a 99\% chance.
And thus, if one sought to exclude the hypothesis of repeatability via a
$p$-value test, even collecting two-orders of magnitude more data only
marginally excludes the hypothesis. Nevertheless, black swans detected in
the early phases of surveys and then found to not repeat would be good
candidates to apply this approach to.

Our work also considers the benefit-to-cost ratio of continuing to
monitor black swan sources. Although valuing scientific discoveries
is somewhat ill-defined, we find that amongst several possible models,
the most conservative models suggest observing for up to 10 times
longer than the initial discovery run is generally expected to
worthwhile. Since an assumption of this work is a fixed observing mode,
one might question the applicability of these results since the survey
mode may be unalterable regardless. However, some surveys, such as 
Breakthrough Listen for example \citep{isaacson:2017}, routinely monitor the
same sample of stars with unchanging sensitivities but the dwell-time on each
target can be easily changed in response to black swan events. As another
example, even if the observing mode is fixed, the effort-level - in particular
the computational resources devoted to a search within a data set -
are resource-limited and thus could be similarly optimized.

Our work provides a starting point for statistically interpreting
black swans, but each event will undoubtedly have its own curiosities
and quirks deserving bespoke attention.

\section*{Acknowledgments}

Thank-you to the reviewer for a helpful report that improved this paper.
Special thanks to donors to the Cool Worlds Lab: Tom Widdowson, Mark Sloan, Douglas Daughaday, Andrew Jones, Jason Allen, Marc Lijoi, Elena West, Tristan Zajonc, Chuck Wolfred, Lasse Skov, Geoff Suter, Max Wallstab, Methven Forbes, Stephen Lee, Zachary Danielson, Vasilen Alexandrov, Chad Souter, Marcus Gillette, Tina Jeffcoat \& Jason Rockett.

\section*{Data Availablity}

No data was used in the preparation of this manuscript.

%\appendix

%

\bsp
\label{lastpage}
\end{document}